\documentclass[twocolumn]{aastex7}

\usepackage{newtxtext}

\usepackage[T1]{fontenc}
\usepackage{ae,aecompl}

\usepackage{graphicx}	
\usepackage{amsmath}	
\usepackage{amssymb}	

\def\gtorder{\mathrel{\raise.3ex\hbox{$>$}\mkern-14mu
    \lower0.6ex\hbox{$\sim$}}}
\def\ltorder{\mathrel{\raise.3ex\hbox{$<$}\mkern-14mu
    \lower0.6ex\hbox{$\sim$}}}

\begin{document}

\title{Direct Collapse Pre-supermassive Black Hole Objects as Ly\,$\alpha$ Emitters}



\author[0000-0002-2243-2790]{Yang~Luo}
\affiliation{Department of Astronomy, and Key Laboratory of Astroparticle Physics of Yunnan Province, Yunnan University, Kunming, Yunnan 650091, China} 
\email[show]{luoyang@ynu.edu.cn (YL)}

\author[0000-0002-1233-445X]{Isaac~Shlosman}
\affiliation{Department of Physics \& Astronomy, University of Kentucky, Lexington, KY 40506-0055, USA} 
\email[show]{isaac.shlosman@uky.edu (IS)}

\shortauthors{Luo and Shlosman}

\begin{abstract}
The Direct Collapse scenario to form the supermassive black hole (SMBH) seeds offers the most promising way to explain the origin of quasars at $z>7$. Assuming atomic primordial gas, can Ly\,$\alpha$ photons escape from the central regions of the collapse and serve as a diagnostic for the detection of these pre-SMBH objects? Models of spherical collapse have found these photons to be trapped and destroyed. We use Ly\,$\alpha$ radiation transfer within the inflow-outflow geometry, based on earlier zoom-in cosmological modeling involving radiation transfer and magnetic forces. Adopting geometry that includes ongoing disk and spherical accretion, and formation of a biconical outflow funnel, we obtain the formation of a dense radiatively driven expanding shell. The Ly\,$\alpha$ transfer is performed using a Monte Carlo algorithm, accounting for the destruction of Ly\,$\alpha$ photons and the emergence of two-photon emission. We find that a substantial fraction of Ly$\,\alpha$ photons can escape through the funnel and calculate the line profiles, the line peak velocity shift, asymmetry, and cuspiness, by varying basic model parameters. The escaping Ly\,$\alpha$ emission is anisotropic and sensitive to the overall inflow-outflow geometry. The escaping fraction of Ly\,$\alpha$ radiation exceeds 95\% from a $z=10$ pre-SMBH object --- in principle detectable by the JWST NIRSpec in the MOS mode, during $\sim 10^4$\,seconds for a $10\sigma$ signal-to-noise ratio. Moreover, comparisons with line shapes from high-$z$ galaxies and quasars allow us to separate them from pre-SMBH objects based on the line shape: the pre-SMBH lines show a profound asymmetry and extended red tail.
\end{abstract}


\keywords{Galaxy formation(595) --- Quasars(1319) --- Supermassive black holes(1663) --- Gravitational collapse(662)  --- Hydrodynamics(1963) --- Radiative processes(2055)}



\section{Introduction}
\label{sec:intro}

Detections of high-redshift quasars now extend to $z > 7$, which is less than 750\,Myr after the Big Bang \citep[e.g.,][]{fan03,mortlock11,willott10,wu15,venemans17,banados18,wang21,bosman24}. The masses of supermassive black holes (SMBHs) that power these objects have been estimated at $\sim 10^9\,M_\odot$, with a plausible discovery of smaller SMBHs $4\times 10^7\,M_\odot$ at $z\sim 10.3$ \citep{bogdan24}, $\sim 10^7\,M_\odot$ at $z\sim 8.7$ \citep{larson23} and $2\times 10^6\,M_\odot$ at $z\sim 10.5$ \citep{maiolino24}. The pathway to forming such SMBHs is still under investigation. The age of the universe makes it very difficult, maybe even unsurpassable, for these SMBHs to form from Population\,III stellar remnants or from stellar mergers in compact clusters. Direct collapse of the gas within dark matter (DM) halos alleviates many problems, but issues remain \citep[e.g.,][]{rees84,haehnelt93,loeb94,bromm03,koushiappas04,begelman06}. An alternative view is that the radiation from these objects is highly beamed, and therefore the masses of their SMBHs are substantially overestimated \citep[e.g.,][]{king24}.

Within the context of Direct Collapse, the model presumes the SMBH growth from much more massive seeds of $\sim 10^5 - 10^7\, M_\odot$ \citep[e.g.,][]{bromm03,wise08,begelman09,regan09,inayoshi20}. Direct collapse occurs when the virial temperature of the parent DM halos becomes higher than the gas temperature. If the temperature of the accreting gas remains at the floor of the atomic hydrogen cooling of $\sim 10^4$\,K \citep[e.g.,][]{greif11,choi13,latif13,choi15,shlosman16,latif16}, the gas fragmentation and the associated star formation will be suppressed. The molecular hydrogen cooling can be suppressed due to background UV radiation, which is also one of the conditions required to avoid fragments \citep[e.g.,][]{dijkstra08,agarwal12,dijkstra14,inayoshi15,maio19,luo20,patrick23,bhowmick22}. 

Accretion disk on scale of $\ltorder 1$\,pc has been shown to form, and the associated problem of angular momentum barrier can be overcome by gravitational torques \citep{choi13,choi15} or by growing magnetic field generated by the magneto-rotational instability, hereafter MRI \citep[][]{chandrasekhar60,balbus91,balbus98}. 

The final phase of the collapse depends on whether the radiation will be trapped or will find an escape route from the center. In the former case, it leads to the formation of a hydrostatic entity \citep{begelman06,begelman08,begelman10}, such as a supermassive star (SMS), in the latter case to a collapsing self-gravitating disk \citep{begelman09}. In both cases, the photosphere formed around this central massive object emits radiation flux near the Eddington limit \citep{luo18,ardaneh18,luo23}. Even if the SMS has formed, the associated accretion disk can form around it \citep{shlosman16}.

Our modeling is based on previous works on this subject \citep{luo18,ardaneh18,luo23}. Direct Collapse scenario supposed to explain the mass accumulation in the center of DM halo, before the SMBH seed has been formed. The accepted pathway is via a supermassive star (SMS), self-gravitating disk, or a mix of these options. The details of this evolution are under investigation, mostly using numerical simulations. However, the only attempt to follow the growth of the central mass accumulation up to few\,$\times 10\,M_\odot$ in the cosmological context and using realistic opacities, including H$^-$, has been performed by \citet{ardaneh18}, although isolated collapse has been modeled by \citet{luo18} and \citet{kimura23}. 

The claimed difference between these models is the spherically-averaged effective temperature of the central mass, namely $\sim 5,000$\,K in \citet{kimura23} versus $\sim 16,000$\,K in \citet{ardaneh18}. However, Kimura et al. have used initial conditions for an isolated, 0.03\,pc spherical Bonner-Ebert sphere \citep{bonnor56} without the background DM halo, which necessarily results in a monotonically increasing accretion rate with time, as observed in \citet{luo18}. Compared to cosmological initial conditions, the isolated collapse misses a number of important points that affect the evolution, becoming non-monotonic is one of them, breaks the spherical and axial symmetry much earlier and more substantially, and shows that the central object is far from being in virial equilibrium. Furthermore, the photosphere has a complex geometry, at least for masses less than few\,$\times 100\,M_\odot$. For example, when the object has reached $10\,M_\odot$, we see no difference between the photosphere size along the equatorial plane of our object \citep{ardaneh18} and that in \citet{kimura23}. Only spherical averaging leads to the difference. Furthermore, in all of our models, a radiatively-driven wind develops, which affects the environment of the growing central object. 

For direct collapse accretion rates, a cold infalling shell can obscure the photosphere of the SMS, but interactions between this shell and outflowing UV radiation are subject to a number of instabilities, e.g., Rayleigh-Taylor instability. Moreover, angular momentum in the accreting matter can lead to various non-axisymmetric instabilities, to amplification of magnetic fields, field expulsion along the spin axis, etc. Under these conditions, it is difficult to avoid dependence on the polar angle, especially keeping in mind that the low angular momentum material will collapse earlier, naturally leading to the formation of funnel along the spin axis. 

In the present work, we avoid the above complications and follow \citet{luo23}, assuming the central mass as a zero-age main sequence (ZAMS) SMS of $10^5\,M_\odot$, whose effective temperature is $6\times 10^4$\,K, emitting close to the Eddington limit.  This $T_{\rm eff}$ generates UV flux which affects the accretion flow, forming the polar outflow channel. Evolution of this outflow and the formation and escape of Ly\,$\alpha$ photons along it are the focus of current work. For comparison, such ZAMS star is expected to have $T_{\rm eff}\sim 10^5$\,K \citep[e.g.,][]{fuller86,begelman10}. Note also that metal-free stars emit more ionizing photons per unit stellar mass \citep[e.g.,][]{tumlinson00}. \citet{hosokawa13} argue that the SMS of this mass in the presence of an unresolved accretion disk has $T_{\rm eff}\sim 10^4$\,K (see their Fig.\,1). They do not refer to the ZAMS star and argue that the object will lie rather on the Hayashi track due to the H$^-$ opacity present in the accretion flow. This argument in \citet{hosokawa13} and \citet{kimura23} is based on a highly idealized subgrid model of accretion and its interaction with the radiation field --- a very complex issue. 

Ly\,$\alpha$ emission has been invoked to detect high-$z$ objects since \citet{partridge67}, specifically the Ly\,$\alpha$ emitting (LAE) galaxies. The observations of LAEs suggest that these objects are the primary population of starforming galaxies or active galactic nuclei at high redshifts \citep[e.g.,][and refs. therein]{ouchi20}. 

These Ly\,$\alpha$ photons are produced during recombination in the gas subjected to the UV continuum. During direct collapse, the Ly$\,\alpha$ emission is one of the prominent cooling processes \citep{cen92,dijkstra16} contributing to the change in the gravitational binding energy. The cooling luminosity is expected to be $L\sim \epsilon\dot M(r) v_{\rm in}^2$, where $\dot M(r)$ is the accretion rate at radius $r$, $v_{\rm in}$ is the radial inflow velocity, and $\epsilon$ is the fraction of accretion energy converted to radiation. After the gas breaks the barrier around $\sim 1$\,pc-scale, the accretion rate and the radial velocity increase along with the gas inflow, until the thermal pressure may start to affect the inflow around $\sim $\,AU-scale \citep{begelman06,begelman09,choi13,luo16,luo18,ardaneh18,luo23}. Hence, the Ly\,$\alpha$ photons can be created in the range of radii and bear a large fraction of the radiation flux in this process. The important question is whether and what fraction of these photons escape from the central region and can be detected.

\citet{ge17} have analyzed the formation and escape of Ly\,$\alpha$ photons using a cosmological simulation of direct collapse, assuming Monte Carlo radiation transfer under uniform density and isotropic emission. As a result, the photon trapping has been maximized inside the central parsec, raising the gas temperature and its Jeans mass. The main effect of these conditions is to delay the initial collapse stage. 

Simulations of direct collapse with radiative transfer for continuum photons or with MHD have modeled collapse in cosmological zoom-in simulations and have shown the formation of accretion disks inside the central 0.1--1\,pc prior to SMBH, associated with a bi-polar outflow \citep{luo23,luo24}. This outflow has been associated with the formation of funnels and dense outflowing shells. The prevailing geometry of the flow allows for the possibility that Ly\,$\alpha$ photons can in principle escape from the central region. In this work we adopt the geometry from Luo et al. modeling, and post-process it by performing Monte Carlo radiation transfer to test the escape of Ly\,$\alpha$ radiation from these pre-SMBH objects. We calculate the shapes of produced Ly\,$\alpha$ emission lines for a range of basic parameters of the flow, and compare them with the observed lines from LAEs.  

This paper focuses on two main issues. (1) Can Ly\,$\alpha$ photons escape from the innermost $\sim 0.01$\,pc of the direct collapse, and can this flux be detected, e.g., by the JWST observations? (2) Is it possible to differentiate between the Ly\,$\alpha$ emission in the direct collapse halo and that of the LAEs? The outline of the paper is as follows: In section\,2, we describe the numerical model used. Results are shown in section\,3, and discussion and conclusions  are provided in section\,4.

\section{The Model}
\label{sec:model}

We use a modified parallel version of the Monte Carlo code by \citet{zheng02}. The code tracks the scattering histories of individual photons, until they either escape or are destructed. It operates on a $3$D grid, with each grid cell including information on gas temperature, composition, velocity, and flow density. Once a photon is generated, a random frequency drawn from the Voigt function and a random propagation direction are assigned to it. The code then follows the photon trajectory and computes its optical depth or distance traveled. With the resonant scattering, photons diffuse both spatially and in frequency space. Additionally, the code calculates the probability of photon escaping along the line-of-sight direction. More details are given in \citet{zheng02}. In the present work we follow the primordial chemistry network described in \citet{luo23}, which used \textsc{grackle}\,3.1 \citep{smith17}.  
  
For the primordial composition gas used here, the dust is absent, and H$_2$ in the outer DM halo is assumed to be destroyed by the UV background. The collapse proceeds isothermally at the cooling floor with $T\sim 7,000$\,K. However, in the central high-density regions, transitions of the form $2p\rightarrow 2s$ followed by the decay to the ground state by emitting two photons must be accounted for, i.e., collisional de-excitation of the $2p$ state. 

This ensures that for each scattering event, there is a finite probability, $p_{\rm dest}$, that a Ly\,$\alpha$ photon is destroyed. In the code, we rescale the scattering redistribution function by a factor of $1-p_{\rm dest}$, where $p_{\rm dest} = n_{\rm p}C_{\rm 2p2s}/(n_{\rm p}C_{\rm 2p2s}+A_{\alpha})$ is the destruction probability. Here, $n_{\rm p}$ denotes the number density of free protons, and $C_{\rm 2p2s} = 1.8\times10^{-4} {\rm cm^3\,s^{-1}}$ is the collisional rate coefficient, and  $A_{\alpha}$ is the Einstein\,A coefficient of the Ly\,$\alpha$ transition.

For a given gas density and a neutral gas fraction, the fraction of escaped photons, $f_{\rm esc}$, after $N_{\rm sca}$ scattering events, is
\begin{equation}
\label{equ:fesc}
	f_{\rm esc} = (1-p_{\rm dest})^{N_{\rm sca}}.
\end{equation}

Statistically, the number of scatterings is a linear function of the optical depth over distance $dl$, $d\tau = n_{\rm HI} \sigma dl$, where $\sigma$ is the cross section at the line center and $n_{\rm HI}$ is the neutral gas number density. Theoretical calculations show that $f_{\rm esc}$ is a function of the free proton density \citep[e.g.,][]{neufeld90}. As gas density and optical depth rise, some Ly\,$\alpha$ photons may be destroyed, as the photons have been reprocessed into the two-photon emission, which we account for.

We adopt the zoom-in cosmological simulation of \citet{luo23} using the Eulerian adaptive mesh refinement (AMR) code Enzo-2.5 \citep{bryan97,norman99,bryan14}. The evolution leads to the central gas accumulation of $10^5\,M_\odot$ within central $\sim 10^{-4}$\,pc, mimicking the SMS and/or accretion disk with effective temperature of $6\times 10^4$\,K. With angular momentum, the gas settles into a rotationally-supported disk around the central mass accumulation, and the continuum radiation flux drives a bi-conical outflow. 

The outflow imparted a large bi-conical cavity or funnel. The swept-up gas in the funnel has been compressed into a thin expanding shell. The interior of the funnel is filled with a hot shocked gas of a very low density. The Ly\,$\alpha$ radiation originates in the expanding shell, and, depending on the column density of spherical accretion exterior to the shell, can in principle escape the DM halo.  

\begin{figure}
\center
\includegraphics[width=0.47\textwidth,angle=0]{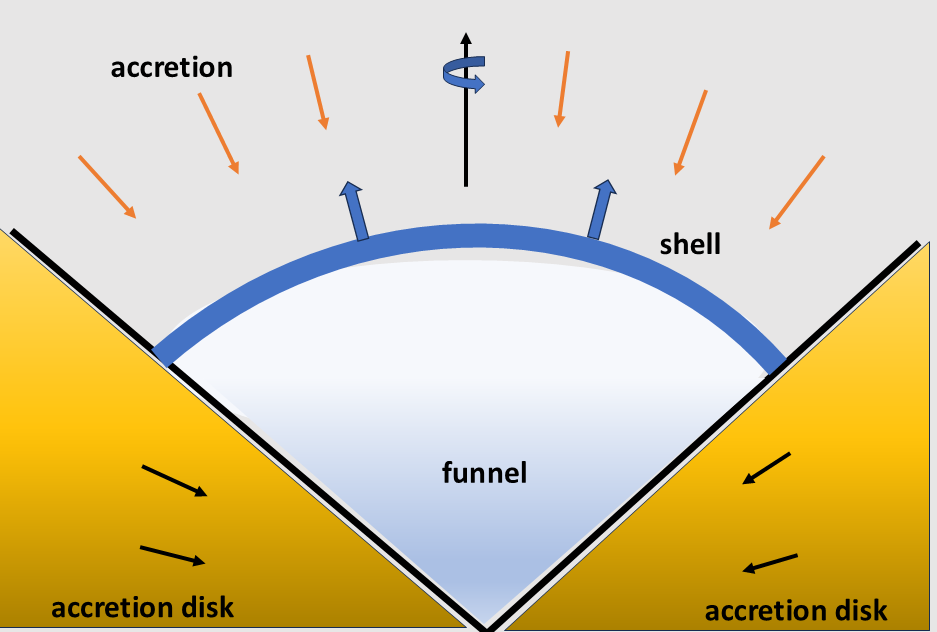}
\caption{The schematic model, not to scale: An edge-on view of the upper hemisphere of disky gas distribution, with a density slice on scale of $2\times 10^{-2}$\,pc, the outflow funnel, radiatively-driven shell, and spherical accretion flow. Arrows indicate the velocity vectors,  with an anisotropic outflow. }
\label{fig:ics}
\end{figure}

Using this geometry adopted from the simulation models, we perform radiation transfer with the Ly\,$\alpha$ photons. The model has four components: the accretion disk flowing toward a central point source, the funnel generated by the outflow, the expanding shell of a compressed gas at $T_{\rm sh} = 8\times10^3$\,K, and the spherical inflow outside this shell with the column density $N_{\rm in}$ (between the shell and $R_{\rm vir}$, as shown schematically in Figure\,\ref{fig:ics}. The range of the model parameters used in modeling is listed in Table\,\ref{tab:para_table}. In the simulated models of \citet{luo23} and \citet{luo24}, the disk is geometrically thick, with the thickness aspect ratio $H_{\rm d}/r_{\rm d} \sim 0.1 - 1$, where $H_{\rm d}$ is the disk thickness, and $r_{\rm d}$ is the disk cylindrical radius. 

In the constructed model, we assume a central source of continuum radiation surrounded by an accretion disk which impacts the expanding shell and generates the Ly\,$\alpha$ luminosity of $10^{43}\,{\rm erg\,s^{-1}}$. The associated outflow above the disk has an opening angle of $\theta_{\rm open}$, forming a low-density hot funnel and a thin dense expanding shell. In the Ly\,$\alpha$ Monte Carlo modeling, the domain box is $2.5\times 10^{-2}$\,pc on each side, and has dimensions of $256^3$. The spatial resolution is $10^{-4}$\,pc. The number of photons launched in each model is $N_{\rm ph} = 10^5$.

\begin{table}
\caption{Model Parameters }
\label{tab:para_table}
\begin{tabular}{lll}
\hline
component  & parameters  & descriptions \\
\hline
spherical inflow	& $N_{\rm in} = 10^{19-22}\,{\rm cm^{-2}}$   & column density	 \\
  & $\tau_{\rm in} = 5.9\times 10^6\ \times  $ & \\
  & $\ \ \ \ N_{\rm in,20}\,(1-x_{\rm in})\,T_{\rm in,4}^{-0.5}$ & optical depth \\
    & $v_{\rm in} = -10\,{\rm km\,s^{-1}}$     &  radial velocity  \\
           & $T_{\rm in} = 10^{4}$\,K     &  temperature    \\
           & $x_{\rm in} = 10^{-2}  $    &  ionization fraction   \\
\hline
disk accretion	   & $\dot{M}_{\rm d} = 1\,{\rm M_\odot\, yr^{-1}}$  & disk accretion rate    \\
           & $v_{\rm d} = -10\,{\rm km\,s^{-1}}$  &  radial velocity   \\
           & $T_{\rm d} = 10^{4}$\,K     &  temperature    \\
           & $x_{\rm d} = 10^{-1}    $  &  ionization fraction   \\
           & $\theta_{\rm open} = 60^\circ-150^\circ  $          & disk opening angel  \\ 
           & $i_{\rm d} = 0^\circ-90^\circ$                  & disk inclination  \\
\hline
outflow funnel	   & $n_{\rm fn} = 100\,{\rm cm^{-3}} $  & funnel density   \\
             & $v_{\rm fn} = 1,000\,{\rm km\,s^{-1}}$ &  radial velocity   \\
           & $T_{\rm fn} = 10^{6}$\,{\rm K}   &  temperature  \\
           & $x_{\rm fn} = 1.0     $   &  ionization fraction   \\
\hline
expanding shell 	   & $\tau_{\rm sh} = 10^{6-8} $     & optical depth    \\
           & $v_{\rm sh} = 100\,{\rm km\,s^{-1}}$   &  radial velocity    \\
           & $T_{\rm sh} = 8,000\, {\rm K}      $ &  temperature   \\
           & $x_{\rm sh} = 10^{-3}     $ &  ionization fraction   \\

\hline
\end{tabular}
\end{table}

For derived Ly\,$\alpha$ profiles, we apply the line diagnostics from \citet{emmering92}, namely,

 (a). {\it The velocity shift of the peak,} ${\rm \Delta v_{\rm p}}$, we use
\begin{equation}
    {\rm \Delta v_{\rm p}} \equiv \frac{\nu_{\rm 0} - \nu_{\rm p}}{\nu_{\rm 0}}\ c,
\end{equation}
where $\nu_{\rm p}$ is the frequency of the line peak, and $\nu_{\rm 0}$ is the frequency of the line center. Here $c$ is the speed of light, and ${\rm \Delta v_{\rm p}}$ is in units of ${\rm km\,s^{-1}}$. The Ly\,$\alpha$ frequency can be converted to the line shift velocity, $\rm \Delta v = c(\nu_0 - \nu)/\nu_0$. The blueshifted line wing, $\rm \Delta v < 0$,
will be absorbed by the intergalactic medium along the line-of-sight and hence is not shown here.

(b). {\it The half-width of the line at $x$, $W_{\rm x}$,} defining $x$ as a given fraction of the peak flux, such that
\begin{equation}
  x = \frac{F(\rm \Delta v_{\rm x})}{F(\rm \Delta v_{\rm p})},
\end{equation}
where $F(\rm \Delta v)$ is the flux at $\rm \Delta v$. For a single peaked profile, there exist two solutions for $\rm \Delta v_{\rm x}$ --- one on the blue side of the peak, $\rm \Delta v_{\rm bx}$, and one on the red side of the peak, $\rm \Delta v_{\rm rx}$. The quantity $W_{\rm x}$ is defined as
\begin{equation}
    W_{\rm x} \equiv \rm \Delta v_{\rm rx} - \rm \Delta v_{\rm bx}.
\end{equation}

(c), {\it The line asymmetry at $x$, $A_{\rm x}$,} where
\begin{equation}
    A_{\rm x} \equiv \frac{(\rm \Delta v_{\rm bx} - \rm \Delta v_{\rm p})-(\rm \Delta v_{\rm p} - \rm \Delta v_{\rm rx})}{\rm \Delta v_{\rm bx} - \rm \Delta v_{\rm rx}}
\end{equation}
provides a measure of the skewness of the profile.

(d). {\it The line cuspiness at $x$, $C_{\rm x}$,} where
\begin{equation}
    C_{\rm x} \equiv \frac{W_{\rm 2x-1}-2W_{\rm x}}{W_{\rm 2x-1}}.
\end{equation}
Here $C_{\rm x}$ is defined so that for a square profile, $C_{\rm x} = -1$; for a triangular profile, $C_{\rm x} = 0$; and for an extremely leptokurtic\footnote{This is the measure of {\it kurtosis}, i.e., of the combined weight of a distribution’s tail relative to the center of the normal distribution curve (the mean). Leptokurtic means more than normal.} profile, $C_{\rm x} = 1$. In this work, we adopt the value of the cuspiness $C_{\rm 0.5}$, which is $C_{\rm x}$ at $x=0.5$.

\begin{figure}
\center
\includegraphics[width=0.45\textwidth,angle=0]{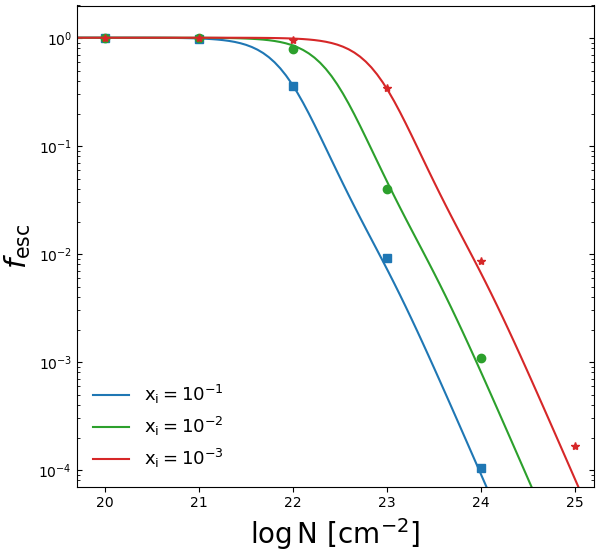}
\caption{The escaping fraction of Ly\,$\alpha$ photons, denoted as $f_{\rm esc}$, are {\it not} destroyed after $N_{\rm scat}$ scattering events. Shown here as a function of the column density $N$ of a spherical inflow. According to Equation\,\ref{equ:fesc}, $f_{\rm esc}$ depends on both the neutral (linked to $N_{\rm scat}$) and ionized gas fractions (linked to $p_{\rm dest}$). Models with varying ionization fractions, $x_{\rm i}$, are differentiated by colors. Solid lines represent the theoretical escape probabilities adopted from \citet{neufeld90}. Our numerical models are denoted by squares, circles, and stars. }
\label{fig:fesc}
\end{figure}

Analytical progress has been made for the emergent Ly\,$\alpha$ spectrum from a {\it static} and optically thick slab or sphere \citep[e.g.,][]{neufeld90,dijkstra06}. The typical Ly\,$\alpha$ flux profile from a static medium is a double-peaked line which is symmetric with respect to the line center \citep{laursen09, dijkstra16}. The location of the peaks is determined by the optical depth. With higher optical depth, the peaks lie farther away from the line center. The line profile also becomes wider with higher optical depth \citep[for a review, see][]{dijkstra17}.

We have tested the code with Ly\,$\alpha$ destruction into two-photon emission, for the case of a spherical accretion, to compare with the analytical solution of \citet{neufeld90}. The column density of spherical accretion, in the range of $N_{\rm HI}\sim 10^{20-24}\,{\rm cm^{-2}}$, has been calculated from a radius $R$ --- the position of the expanding shell, and outwards to $R_{\rm vir}$. With increasing $N_{\rm HI}$, the Ly\,$\alpha$ photons can be destroyed by the two-photon production process. As shown in Figure\,\ref{fig:fesc}, $f_{\rm esc}$ decreases with increasing column density, and the agreement between the analytical and numerical solutions is excellent. 

As a next step, we invoke the velocity field of the medium --- this point is important for the model evaluated here. The outflow funnel enables the continuum photons to reach the expanding shell via region of a low optical depth. An expanding medium can cause suppression of the blue wing of the line and enhancement of the red wing. Conversely, a collapsing medium exhibits an enhanced blue wing and a suppressed red wing. The position of the peaks in this case is also determined by the medium velocities \citep{zheng02,laursen09}.  

\section{Results}
\label{sec:results}

\begin{figure*}
\center
\includegraphics[width=0.8\textwidth,angle=0]{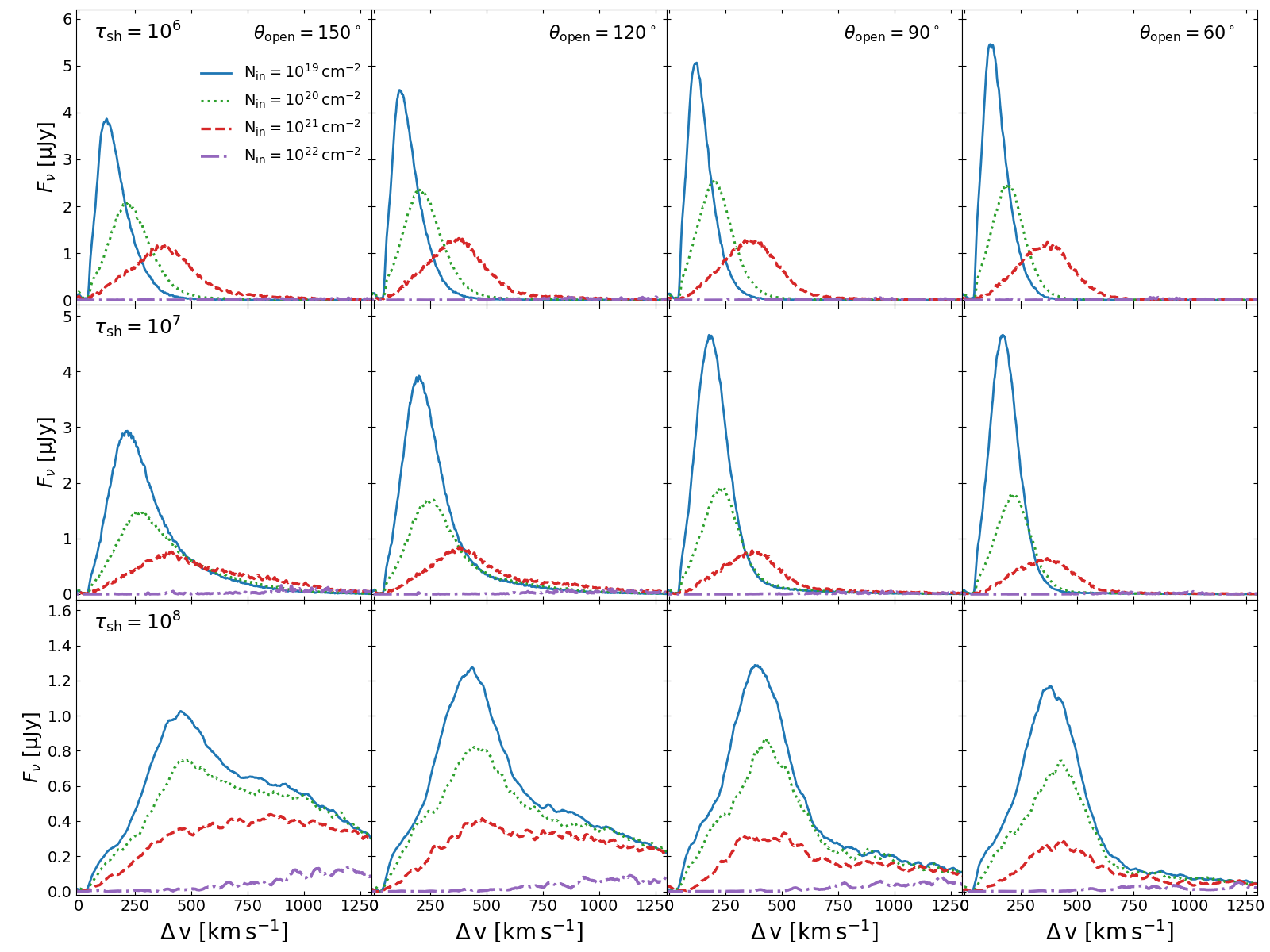}
\caption{The set of resulting Ly\,$\alpha$ line profiles in our models, which display the Ly\,$\alpha$ radiation flux as a function of the velocity offset of the emission line from the systemic redshift, $\Delta v$. The photons have been injected at the line center with $T=10^4$\,K. Horizontal frames: varying the funnel opening angles, $\theta_{\rm open} = 150^\circ - 60^\circ$. Vertical columns: varying the optical depth of the spherical accretion outside the shell, $\tau_{\rm sh} = 10^6-10^8$ (vertically). Within each frame, dependence of the line on the column density of the spherical accretion outside the shell has been varied in the range of $N_{\rm in} = 10^{19}-10^{22}\,{\rm cm^{-2}}$.
}
\label{fig:models}
\end{figure*}

We consider the source to be at the redshift of $z = 10$, with the Ly\,$\alpha$ luminosity of $L = 10^{43}\,\rm erg\,s^{-1}$. The resulting flux is  
\begin{equation}
F = L/4\pi D^2 \approx 7.4\times10^{-18}\,\rm erg\,s^{-1}\,cm^{-2}\ L_{43} D_{10}^{-2},
\label{eq:flux}
\end{equation}
with $L_{43}$ in units of $10^{43}\,\rm erg\,s^{-1}$, and $D_{10} \approx 1.06\times10^5\,\rm Mpc$ being the luminosity distance at $z = 10$, utilizing the cosmological parameters from the Planck 2015 data \citep{planck-collaboration16} as follows: $\Omega_{\rm m} = 0.3089$, $\Omega_\Lambda = 0.6911$, $\Omega_{\rm b} = 0.04859$, $\sigma_8 = 0.8159$, $n_{\rm s} = 0.9667$, and the Hubble constant $h = 0.6774$ measured in units of $100\,\rm km\,s^{-1}\,Mpc^{-1}$. The Ly\,$\alpha$ photons are redshifted to $\lambda\sim 1.34\,\mu$m for the observer.
 
We run models with all four components, using the fiducial values provided in Table\,\ref{tab:para_table},  by varying the following parameters within the range shown in the table: the column density of the spherical accretion $\rm N_{\rm in}$, the optical depth of the expanding shell $\tau_{\rm sh}$, and opening angle of the accretion disk $\theta_{\rm open}$ (i.e., the width of the funnel). The remaining parameters are sourced from the simulation models in \citet{luo23} and kept constant during the calculations. While these remaining parameters have the potential to slightly modify the spectral line profiles, their effects are disregarded in our current calculations. When the cavity is cleared by the outflow, the gas number density within can decrease to around $\sim 100\,\rm cm^{-3}$, with the ionization fraction nearly one, rendering the Ly\,$\alpha$ optical depth insignificant. The resulting line profiles are double peaked, but only the enhanced red wing of the lines is shown here. 

\begin{figure*}
\center
\includegraphics[width=0.99\textwidth,angle=0]{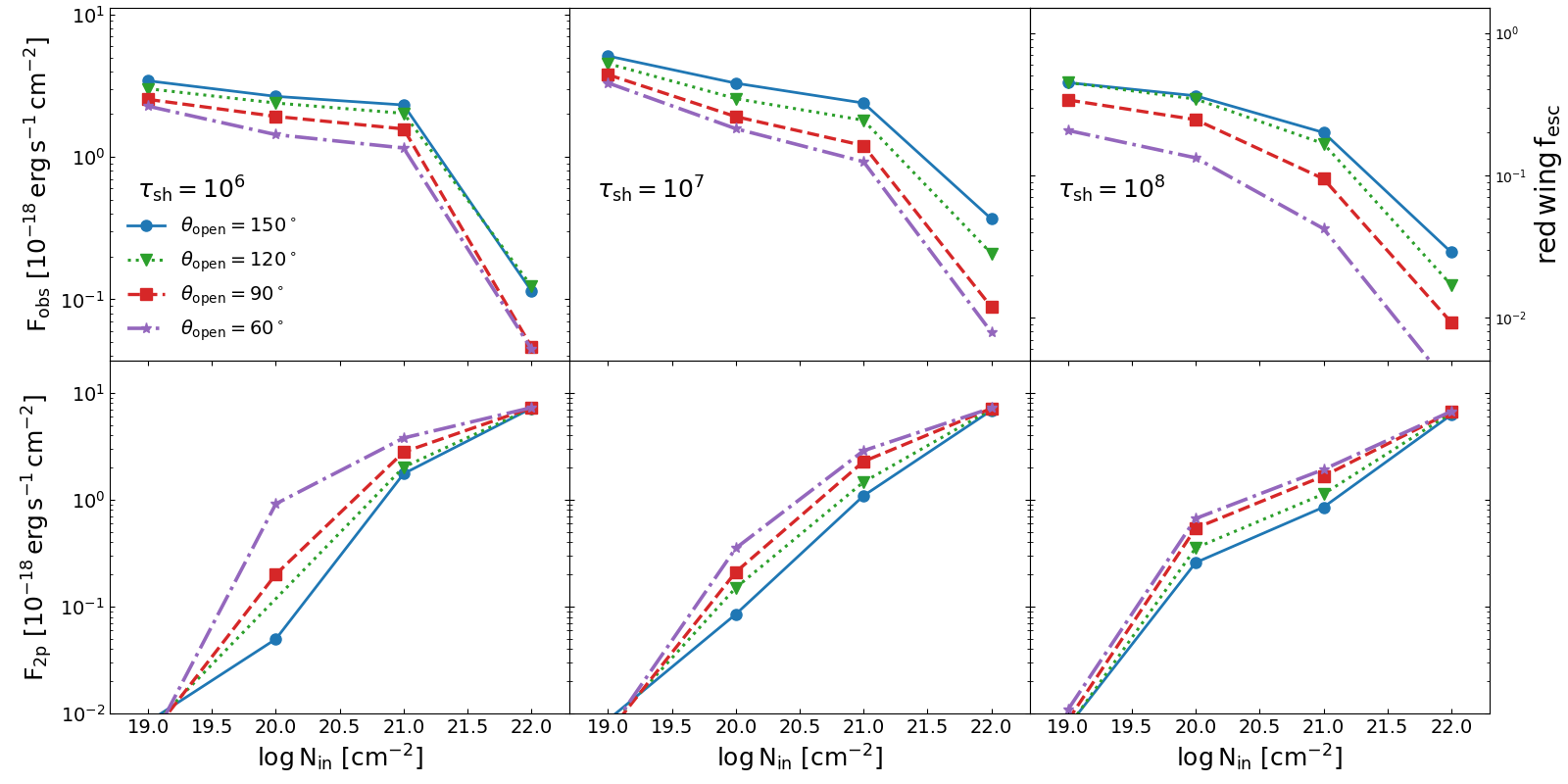}
\caption{Top row: The observed integrated flux of Ly\,$\alpha$ photons (left $y$-axis), as a function of $N_{\rm in}$ for $\tau_{\rm sh} = 10^6-10^8$ and $\theta_{\rm open} = 60^\circ-150^\circ$. Right $y$-axis gives the fraction of the observed red wing flux in units of the total intrinsic flux. Note that only the red wing is detectable. Bottom row: The integrated  two-photon flux as a function of the same variables.}
\label{fig:fluxes}
\end{figure*}

\begin{figure*}
\center
\includegraphics[width=0.99\textwidth,angle=0]{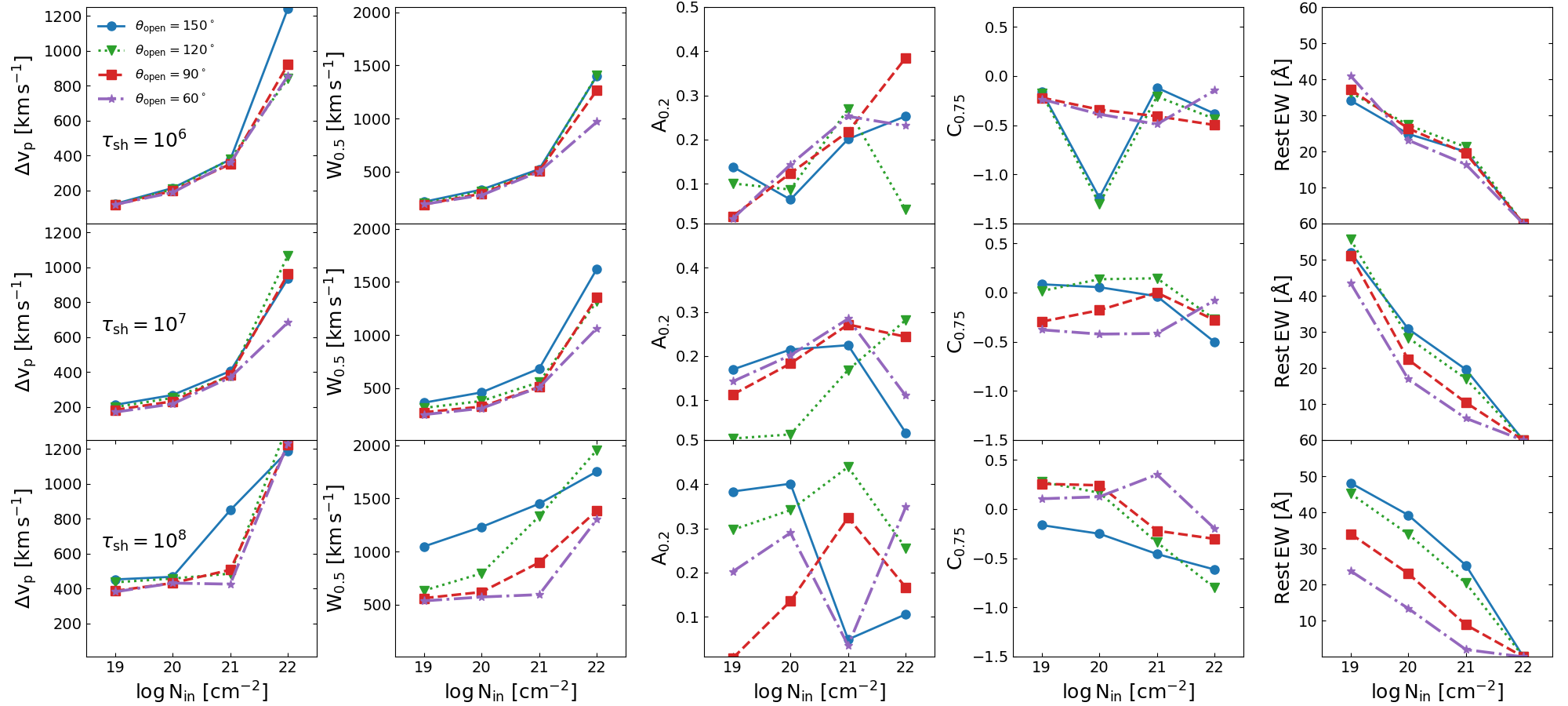}
\caption{The Ly\,$\alpha$ line shape diagnostics defined in section\,\ref{sec:model} as a function of the inflow column density $N_{\rm in}$ for $\tau_{\rm sh} = 10^{6-8}$. From left to right: $\Delta v_{\rm p}$ --- the red wing peak velocity shift; $W_{0.5}$ --- half-width at $x=0.5$;  $A_{0.2}$ --- the asymmetry at $x=0.2$; $C_{0.75}$ --- the cuspiness at $x=0.75$; the equivalent width, EW, in the rest frame. 
}
\label{fig:diagnostics}
\end{figure*}

The initial model has been run assuming only a coexistence of the disk and spherical accretion, in the presence of the point source of the UV continuum at the center and a weak outflow. This model corresponds to the early stages of central mass accumulation in \citet{luo23}. As expected, the emerging Ly\,$\alpha$ photons have been mostly attenuated by the two-photon process already in the central region, and the emerging flux has been well below the JWST sensitivity.  

At later epochs, \citet{luo23} observed the formation of a funnel along the rotation axis of the accretion disk, formation of an expanding shell, and reduction of the density in the funnel. In this regime, a strong flux of Ly\,$\alpha$ photons has emerged. We varied the properties of the expanding shell, column density of the spherical accretion, and the opening angle of the funnel to analyze the dependence of the emerging flux on these values. Figure\,\ref{fig:models} provides the set of emerging Ly\,$\alpha$ line profiles and fluxes formed under these conditions. We analyze them by modifying one parameter at the time.

As shown in Figure\,\ref{fig:models}, the peak Ly\,$\alpha$ fluxes are inversely proportional to the column densities of the spherical accretion $N_{\rm in}$. For $N_{\rm in}\sim 10^{22}\,{\rm cm^{-2}}$, the emerging flux is only $\sim 1\%$ compared to that for $N_{\rm in}\sim 10^{19}\,{\rm cm^{-2}}$. The red peak of the line shifts gradually to larger $\Delta {\rm v}$ with increasing $\tau_{\rm sh}$. The line full width at half maximum (FWHM), for models $N_{\rm in}\ltorder 10^{21}\,{\rm cm^{-2}}$, increases with $\tau_{\rm sh}$, e.g., from $\sim 200\,{\rm km\,s^{-1}}$ to $\sim 1300\,{\rm km\,s^{-1}}$ for $\tau_{\rm sh}\sim 10^6$ increase to $10^8$.

The location of the red peak of Ly\,$\alpha$ line is sensitive with respect to all three basic parameters used. It increases monotonically with increasing $N_{\rm in}$ and with increasing $\tau_{\rm in}$. Dependence on $\theta_{\rm open}$ is not so significant. It slightly increases with increase in $\theta_{\rm open}$, for $N_{\rm in}\sim 10^{22}\,{\rm cm^{-2}}$, but for $N_{\rm in}\ltorder 10^{20}\,{\rm cm^{-2}}$, it is largely indifferent to $\theta_{\rm open}$. 

To assess the Ly\,$\alpha$ detectability, we use the Ly\,$\alpha$ source at $z=10$ and apply the JWST NIRSpec at appropriate 1.34\,$\mu$m. For NIRSpec in the multi-object spectroscopy (MOS) mode, the sensitivity is approximately $\sim 0.1\,\mu\rm{Jy}$, assuming a $10\sigma$ signal-to-noise ratio over $10^4$ seconds of exposure\footnote{https://www.stsci.edu/jwst/science-planning/proposal-planning-toolbox/sensitivity-and-saturation-limits}. As illustrated by Figures\, \ref{fig:models} and also \ref{fig:fluxes}, the calculated flux density for the Ly\,$\alpha$ line from our source exceeds this sensitivity level, especially for model with $N_{\rm in}\ltorder 10^{21}\,{\rm cm^{-2}}$. 

Figure\,\ref{fig:fluxes} (top row) shows the frequency-integrated Ly\,$\alpha$ flux, $F_{\rm obs}$, as a function of $N_{\rm in}$, $\tau_{\rm sh}$, and $\theta_{\rm open}$. The integrated Ly\,$\alpha$ flux appears flat for $N_{\rm in}\sim 10^{19}-10^{21}\,{\rm cm ^{-2}}$, followed by a sharp decline at higher column density by a factor of $\sim 30$.  For higher $\tau_{\rm sh}$, the behavior is similar, with the lines spreading out with increasing $N_{\rm in}$. The flux is basically independent of the funnel opening angle; the decline becomes shallower for larger funnel opening angles.  

The bottom row of Figure\, \ref{fig:fluxes} displays the integrated two-photon flux, $F_{\rm 2p}$, as a function of the same variables. This flux increases with $N_{\rm in}$ by about 3 orders of magnitude, and appears to be nearly independent of $\tau_{\rm sh}$ and the funnel opening angle.  

Figure\,\ref{fig:diagnostics} provides the line diagnostics for all models shown in Figure\,\ref{fig:models}. The left column displays the $\Delta{\rm v_{\rm p}}$ -- the red peak velocity shift as a function of the spherical accretion column density, $N_{\rm in}$, for a number of funnel opening angles and the optical depth of the expanding shell. The peak velocity shift $\rm \Delta v_{\rm p}$ increases with $N_{\rm in}$, and this shift increases faster for $N_{\rm in}\gtorder 10^{21}\,{\rm cm^{-2}}$, from $\Delta v_{\rm p}\sim 150-400\,{\rm km\,s^{-1}}$ to $\sim 1,200\,{\rm km\,s^{-1}}$. Moreover, above $N_{\rm in}\sim 10^{21}\,{\rm cm^{-2}}$, the shift increases faster for larger $\theta_{\rm open}$, for $\tau_{\rm sh}\sim 10^6-10^7$. For higher $\tau_{\rm sh}$, this dependence is stronger and enters at lower $N_{\rm in}$.

The second column in Figure\,\ref{fig:diagnostics} provides $W_{0.5}$ --- the FWHM at $x=0.5$. For all cases, the FWHM increases with $N_{\rm in}$, and $W_{0.5}(N_{\rm in})$ becomes segregated at higher $\tau_{\rm sh}$. The velocity spread of the FWHM increases from  $\sim 200-500\,{\rm km\,s^{-1}}$ to $\sim 1,000-2,000\,{\rm km\,s^{-1}}$.

The third column in this Figure displays the asymmetry of the line, $A_{0.2}$, at $x=0.2$. Here the dependence on $N_{\rm in}$ is more complex. The asymmetry increases with $N_{\rm in}$ until $10^{21}\,{\rm cm^{-2}}$ for $\tau_{\rm sh}\sim 10^6-10^7$ range, and subsequently decreases. For $\tau_{\rm sh}\sim 10^8$, $A_{0.2}$ behaves in the opposite way for small and large opening angles.

The fourth column shows the cuspiness of the line, $C_{0.75}$, at $x=0.75$. For $\tau_{\rm sh}\sim 10^6-10^8$, $C_{0.75}$ is insensitive to $N_{\rm in}$, with a very small decline with the column density, and a spread by a factor of 3. 

The last column in  Figure\,\ref{fig:diagnostics} displays the rest frame $z=10$ equivalent width, $EW$. To calculate the EW at $z=0$, we use
\begin{equation}
  EW = \int{\bigg(\frac{F_\lambda}{F_{\rm c}} - 1\bigg) d\lambda},  
\end{equation}
where $F_{\rm c}$ is the continuum flux density near the wavelength of LY\,$\alpha$. In the rest frame of the pre-SMBH object, the EW must be divided by $z + 1$, and the obtained EWs lie in the range $\sim 0-60\,\text{\AA}$. For brighter $F_{\rm c}$, $EW$ becomes smaller and vice versa.  

The main contribution to $F_{\rm c}$ is expected to come from the mixture of Pop\,III and Pop\,II stars populating the parent halo. So far, we have accounted for their effect on Direct Collapse only indirectly, by assuming that these stars are responsible for ionizing the gas and preventing the formation of H$_2$. To obtain $F_{\rm c}$, we must estimate the blackbody flux from these stars. We assume the total stellar luminosity of a primeval galaxy in the early stage of Direct Collapse lies at $\sim 10^{38}\,\rm erg\,s^{-1}$, and the effective temperature of their blackbody is $\sim 10^4$\,K. The resulting $F_{\rm c}\sim 1.2\times10^{-20}\,{\rm erg\,cm^{-2}\,s^{-1}}\,\text{\AA}^{-1}$. In our calculations, for simplicity, we assume $F_{\rm c} = 10^{-20}\, \rm erg\,s^{-1}\,cm^{-2}\,\text{\AA}^{-1}$.

The continuum flux extracted from the synthetic spectra of simulated first galaxies has been estimated to be around $\sim 8\times10^{-21}\,{\rm erg\,cm^{-2}\,s^{-1}}\,\text{\AA}^{-1}$ \citep[e.g.,][]{zackrisson17}. Moreover, recent JWST observations of high-$z$ quasars at $z > 8$, similarly give the continuum flux around $F_{\rm c}\sim {\rm few}\times 10^{-20}\, \rm erg\,s^{-1}\,cm^{-2}\,\text{\AA}^{-1}$ \citep[e.g.,][]{bunker23,larson23}.  

As the inflow column density $N_{\rm in}$ increases, the rest EWs generally decline. When $N_{\rm in}$ exceeds $10^{22}\,{\rm cm^{-2}}$, the EWs approach zero. Comparing the first and last columns of Figure\,\ref{fig:diagnostics}, we see that with an increase in inflow column density, $\Delta\,{\rm v}_{\rm p}$ increases while the rest EWs decrease, indicating a negative correlation between $\Delta\,{\rm v}_{\rm p}$ and the EWs.

\begin{figure} 
\center
\includegraphics[width=0.48\textwidth,angle=0]{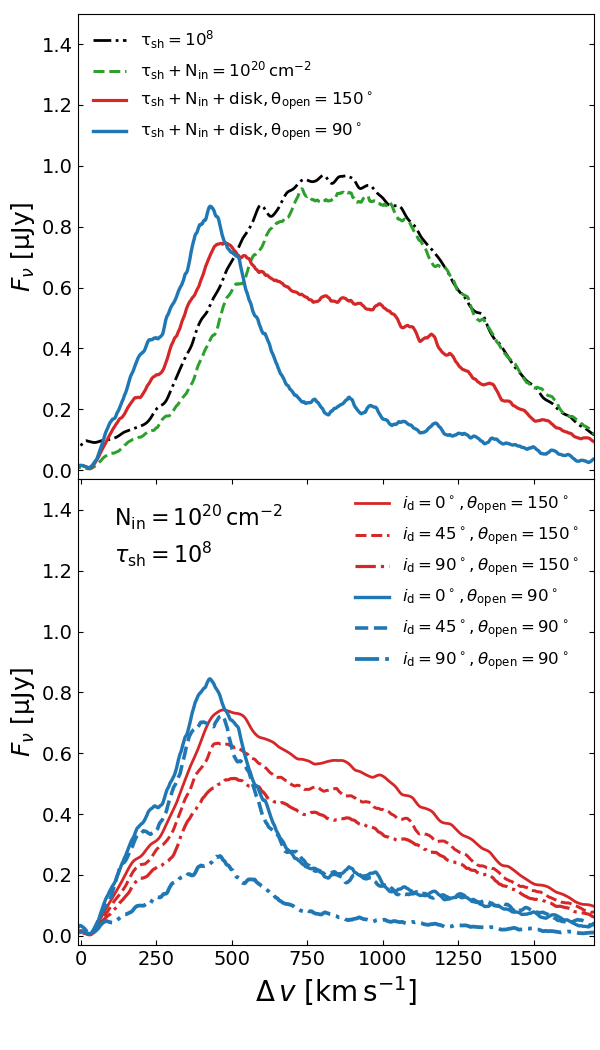}
\caption{Top panel: line profiles shown in Figure\,\ref{fig:models} for $\tau_{\rm sh} = 10^8$, but (a) no accretion disk and no spherical accretion, with $\theta_{\rm open}= 180^\circ$ (black dot-dashed line; (b) no accretion disk, $N_{\rm in} = 10^{20}\,{\rm cm^{-2}}$, $\theta_{\rm open}= 180^\circ$ (green dashed line); (c) with perfectly absorbing accretion disk, $N_{\rm in} = 10^{20}\,{\rm cm^{-2}}$, $\theta_{\rm open}= 150^\circ$ (red solid line); (d) with perfectly absorbing accretion disk, $N_{\rm in} = 10^{20}\,{\rm cm^{-2}}$, $\theta_{\rm open}= 90^\circ$ (blue solid line). Bottom panel: same as above, but various disk inclination angles, $i_{\rm d} = 0^\circ - 90^\circ$, and the funnel opening angles.}
\label{fig:nodisk}
\end{figure}

Next, we quantify the role of the accretion disk in the escape of the Ly\,$\alpha$ photons, and its effect on the line shape. Figure\,\ref{fig:nodisk} (top frame) displays the line profiles shown in Figure\,\ref{fig:models}, with several modifications to emphasize the contribution of the accretion disk. For comparison, the black dot-dashed line shows the test case when both the disk and the spherical accretion are absent. Only the expanding shell is present, with $\tau_{\rm sh} = 10^8$. In this case $\theta_{\rm open}= 180^\circ$, and the Ly\,$\alpha$ line has a nearly flat top in the range of $\Delta\rm{v}\sim 600-1,100\,{\rm km\,s^{-1}}$. The green dashed line, representing the same $\tau_{\rm sh}$ and $\theta_{\rm open}$, without the accretion disk, but accounting for a spherical accretion with $N_{\rm in} = 10^{20}\,{\rm cm^{-2}}$, closely follows the previous case. The red solid line, for the same $\tau_{\rm sh}$ and $N_{\rm in}$, and for a perfectly absorbing Ly\,$\alpha$ photons accretion disk with $\theta_{\rm open} = 150^\circ$, has the red peak shifted left to $\Delta {\rm v}_{\rm p}\sim 420\,{\rm km\,s^{-1}}$, and $F_{\rm \nu}$ reduced by $25\%$. This line is cuspy and profoundly asymmetric. Finally, the blue solid line with the same parameters but with $\theta=90^\circ$, has a triangular shape with a long high velocity (red) tail. This peak is redshifted to $\Delta {\rm v}_{\rm p}\sim 370\,{\rm km\,s^{-1}}$.

To further analyze the effect of the disk on the line shape, we tracked the photon trajectories, and checked the number of photon scatterings across each component. For the no-disk/no-spherical accretion model shown in Figure\,\ref{fig:nodisk} (top frame, the black dot-dashed line) with $\tau_{\rm sh} = 10^8$, we found the photon escape fraction to be $f_{\rm esc}\sim 0.98$. For the model represented by the green dashed line with $\tau_{\rm sh}$ and $N_{\rm in}$, $f_{\rm esc}\sim 0.95$. In this case, the expanding shell dominates both the optical depth and the number of scatterings, explaining why the line shapes appear similar for these two models.  

With the introduction of the disk component, we have found that the number of scattering increases. However, the fraction of scatterings which occurred in the disk is $\sim 0.3\%$ for $\theta_{\rm open} = 150^\circ$ model, and $\sim 0.6\%$ for $\theta_{\rm open} = 90^\circ$ model. While the disk contribution to the number of scatterings is small, only $\sim 27\%$ of the Ly\,$\alpha$ photons entering the disk manage to escape from it for $\theta_{\rm open} = 150^\circ$ model, and only $\sim 3\%$ escape from the disk for $\theta_{\rm open} = 90^\circ$ model.

This also means that the Ly\,$\alpha$ photons which entered the disk are more easily destroyed with decreasing  $\theta_{\rm open}$, i.e., with decreasing opening angle of the funnel. For $\theta_{\rm open} = 90^\circ$, the disk is essentially a purely absorbing component. The presence of an accretion disk has a substantial effect on all the parameters of the line shape, reducing $\Delta {\rm v}_{\rm p}$ by a factor of 2 and resulting in a red asymmetry of the line, with a long red tail. After being scattered within the disk, the shape of the line undergoes significant modification. As more photons are scattered and absorbed within the disk, the line width and its asymmetry increase significantly, while the velocity of the line peak is slightly reduced. 

The lower frame of Figure\,\ref{fig:nodisk} displays the effect of the observer's inclination angle with respect to the disk polar angle on the line shape. To demonstrate this, we use two opening angles for the funnel, $\theta_{\rm open}= 90^\circ$ and 150$^\circ$. For each of these angles we provide three inclination angles, $i_{\rm d}=0^\circ$, 45$^\circ$, and 90$^\circ$. For $\theta_{\rm open}=90^\circ$, we obtain strongly leptocurtic lines with an elongated red tail. The line peak has shifted slightly to a smaller $\Delta {\rm v}_{\rm p}$. At the largest inclination, $i_{\rm d}=90^\circ$, the peak flux has been reduced by a factor of 4. 

For a wide-opened funnel with $\theta_{\rm open}=150^\circ$, the red tail flux is substantially higher than for the narrower funnel, and the line width $W_{\rm 0.5}$ is also broader. As the inclination varies, the locations of the line peaks for both opening angles mostly remain stable; however, the peak flux decreases as inclination increases. With higher inclinations, the line width $W_{\rm 0.5}$ and asymmetry $A_{\rm 0.2}$ generally become larger.

\section{Discussion and Concluding Remarks}
\label{sec:discuss}

In the context of this work, three issues stand out. First, the ability of Ly\,$\alpha$ photons to escape during the Direct Collapse stage, i.e., prior to the formation of the SMBH seed. To determine this, we based our modeling on the inflow-outflow geometry, which has been justified by previous work involving the radiation transfer and the MHD in zoom-in cosmological simulations \citep{luo23,luo24}. The second issue is the detectability of the Ly\,$\alpha$ flux during the Direct Collapse, e.g., by the NIRSpec of the JWST. Finally, can the calculated Ly\,$\alpha$ line profiles be distinguished from other high-$z$ sources, e.g., LAEs and quasars?

To answer these questions, we refer to Direct Collapse in the pre-SMBH stage at $z=10$, which has accumulated $\sim 10^5\,M_\odot$ within the central $\sim 10^{-4}$\,pc. This mass accumulation surrounded by an accretion disk or SMS/accretion disk and accreting at $\sim 1\,M_\odot\,{\rm yr^{-1}}$ generates a blackbody UV continuum and has developed a bi-conical funnel along its rotation axis. The funnel has been nearly emptied by the radiation force that formed a series of expanding shells, and has triggered the Ly\,$\alpha$ flux of $\sim 10^{43}\,{\rm erg\,s^{-1}}$ from the shell. 

Within the Direct Collapse scenario, cooling of the collapsing gas is dominated by atomic processes\footnote{This assumes implicitly the presence of UV background from stars or else to prevent formation of H$_2$.}, with $\sim 40\%$ attributed to Ly\,$\alpha$ cooling \citep[e.g.,][]{dijkstra14}. As a result, the Ly\,$\alpha$ emission is anticipated to be the dominant spectral feature, based on theoretical predictions. However, near the halo center, where the Ly\,$\alpha$ is most intense, the increased density results in a higher optical depth, significantly hindering the escape of Ly\,$\alpha$ photons. Our modeling shows that when the gas column density between the expanding shell and the virial radius is less than $10^{22}\,{\rm cm^{-2}}$, above 95\% of Ly\,$\alpha$ photons escape, and when $N_{\rm in}$ is $\gtorder 10^{22}-10^{23}\,{\rm cm^{-2}}$, only $1-10\%$ of the radiation manages to escape through the bi-conical funnel. The rest disintegrate into the two-photon emission. This escape is in a sharp contrast with the spherical collapse, where Ly\,$\alpha$ becomes fully trapped, temporarily slowing down the collapse \citep{ge17}. We therefore propose that outflows generated in the presence of the central accretion disk, driven by radiation \citep{luo23} or magnetic forces \citep{luo24}, could clear the gas within the bi-conical funnel, and ultimately reduce the Ly\,$\alpha$ optical depth. We have shown that the escaping Ly\,$\alpha$ flux can be detected by the JWST.

Using the Monte Carlo technique, we have calculated the escaping Ly\,$\alpha$ flux from this pre-SMBH object, including the underlying accretion disk and spherical accretion outside the expanding shell. We have also included the destruction mechanism of these photons, the so-called two-photon channel. We find that for reasonable column densities of spherical accretion in the range of $10^{19}-10^{21}\,{\rm cm^{-2}}$, the emerging flux of a few $\mu$Jy can be detected by JWST with NIRSpec in the MOS mode, reaching a $\sim 10\sigma$ signal-to-noise ratio during $10^4$\,seconds of exposure.

We have obtained the Ly\,$\alpha$ line shapes for a range of model parameters, such as the inflow column densities, funnel opening angles, and optical depth to Ly\,$\alpha$ photon scatterings in the expanding shell. The transfer of Ly\,$\alpha$ photons through the gas medium is notably influenced by the gas distribution, composition, and kinematics. While factors such as gas ionization, velocity, and temperature may be significant, our modeling assumes these parameters within a reasonable range guided by simulations, and does not focus on their variations.

The emerging Ly\,$\alpha$ photons display a double-peaked profile, but the intergalactic medium absorbs the blue wing, leaving only the red wing observable. We find that the peak position of the red wing is influenced by the inflow column density. As the column density increases, both the peak velocity and line width increase, while the peak flux decreases. When the column density reaches $\gtorder 10^{22}\,\rm{cm}^{-2}$, a mere $\sim 10\%$ of the total photons are able to escape. 

We also observe changes in line profiles with an increase in the optical depth of the expanding shell. For the fixed inflow column density, the line broadens when the shell optical depth increases. At the optical depth of $\tau_{\rm sh}\sim 10^8$, corresponding to the shell column density of $\sim 10^{21}\,{\rm cm^{-2}}$, the escape fraction is reduced to $\sim 0.3$. 

Furthermore, we have estimated the effect of the accretion disk, whose column density generally exceeds $\sim 10^{22}\,{\rm cm^{-2}}$, hinting at the possibility that photons being scattered toward the disk might alter the line profiles. By varying the funnel opening angle (i.e., the disk opening angle), the likelihood of photons being scattered into the disk is affected. Our findings indicate that when $N_{\rm in} \ltorder 10^{21}\,{\rm cm^{-2}}$ and $\tau_{\rm sh}\ltorder 10^7$, the funnel opening angle has a minimal effect on the line profiles. Conversely, with greater $N_{\rm in}$ and $\tau_{\rm sh}$ values, the disk can alter the line shapes due to an increased number of scattered photons, which enhances the possibility of them entering the disk and being destroyed.

A characteristic feature of the obtained Ly\,$\alpha$ line profile is the appearance of the extended red tail. For example, the lower left panel of Figure\,\ref{fig:models} displays this tail for $N_{\rm in}\sim 10^{19} - 10^{20}\,{\rm cm^{-2}}$ (represented by the blue and green lines), where for $\Delta {\rm v}_{\rm p}\sim 430\,{\rm km\,s^{-1}}$. This feature can also appear due to the line asymmetry, particularly when $A_{\rm 0.2}\gtorder 0.3$. The extended red tail can also be the result of the line being reshaped by the accretion disk. The extended tail becomes more pronounced with higher $\tau_{\rm sh}$, while it diminishes with lower $\theta_{\rm open}$. The extended red tail can serve as an indicator of the Ly\,$\alpha$ flux coming from the Direct Collapse pre-SMBH object.

Can the Ly\,$\alpha$ emission from LAEs and direct collapse pre-SMBH objects be differentiated?  Statistically, the emergent Ly\,$\alpha$ flux from LAEs is primarily connected to massive stars within \ion{H}{2} regions \citep[e.g.,][]{garel12,yamada12,verhamme17}. The LAEs are observed by a rest-frame Ly\,$\alpha$ line $EW \gtorder 20\,\text{\AA}$\citep[e.g.,][]{ouchi20}. Although our pre-SMBH objects range in $EW\sim 0-60\,\text{\AA}$,
the statistics are not available. Yet, the detection of sources with similar EWs to that in LAEs, in tandem with additional properties, as discussed above, e.g., the appearance of an extended red tail, can point to a population of pre-SMBH objects.
 
The Ly\,$\alpha$ line from LAEs exhibits diverse profiles, featuring single or double emission peaks \citep{yamada12}. The NIR spectroscopic studies of LAEs reveal significant gas outflows, with velocities around $\sim 200\,\mathrm{km\,s}^{-1}$, which generally align with the LAE velocity offset $\Delta {\rm v}_{\rm p}$ \citep[e.g.,][]{ouchi20}. For these LAEs, a negative relationship has been identified between the velocity offset $\Delta {\rm v}_{\rm p}$ and the line EW \citep{hashimoto13,shibuya14,erb14,nakajima18}. A similar trend can be identified in our models from Figure\,\ref{fig:diagnostics}, regarding $\Delta\,{\rm v}_{\rm p}$ versus the line EWs. 

Moreover, recent JWST observations of high-$z$ quasars with $z\gtorder 7$ have identified many Ly\,$\alpha$ emission lines \citep{banados18,wang21,yang21,yang23b,yang23,matthee24,onorato24}. For example, the redshifted line in $\rm GN-z11$, observed at $z\sim 10.6$, is described by a Gaussian profile with minimal asymmetry \citep{bunker23}. The velocity of the line shift is approximately $\sim 555\,\mathrm{km\,s}^{-1}$, and its width is around $\sim 566\,\mathrm{km\,s}^{-1}$. The equivalent width of the line is $\sim 18\,\text{\AA}$.  The Ly\,$\alpha$ line in CEERS\,1019, at $z \sim 8.7$, presents similar characteristics, with a line shift velocity, line width, and equivalent width of approximately $\sim 218\,\mathrm{km\,s^{-1}}$, $\sim 702\,\mathrm{km\,s}^{-1}$, and $\sim 10.1 \text{\AA}$, respectively \citep{larson23}. These velocity offsets and equivalent widths in the rest-frame are comparable to those in luminous galaxies with $7.5 < z < 9$ that show Ly\,$\alpha$ emission. 

To summarize, (1) the Ly\,$\alpha$ photons can escape from the inner region of Direct Collapse halos due to the anisotropic gas distribution. We show that the pre-SMBH direct collapse objects are detectable by JWST NIRSpec, with a reasonable exposure time. 

(2) The Ly\,$\alpha$ line profiles exhibit measurable differences compared to the lines from LAEs and high-$z$ quasars, thus providing an additional channel to study the early stages of SMBH seed formation. A notable characteristic of Ly\,$\alpha$ profiles from the pre-SMBH objects is a significant asymmetry. With increasing optical depth in accretion flow, asymmetry rises while the line cuspiness diminishes. The line exhibits an extended red tail. 

(3) Our Ly\,$\alpha$ EWs show a negative trend with the velocity offset of the red peak, in tandem with a similar effect observed in LAEs and high-$z$ quasars.

Finally, the high-$z$ quasars and LAEs display metal enrichment \citep{yang21,zou24}, unlike the metal-free Direct Collapse pre-SMBH objects. Thus, future observations of the Direct Collapse pre-SMBH objects should focus on searching for metal-free targets with pronounced Ly\,$\alpha$ line asymmetry.


\begin{acknowledgments}
We thank Zheng Zheng for providing us with the Monte Carlo code, as detailed in \citet{zheng02}. All the analysis has been conducted using yt \citep{turk11}, http://yt-project.org/. I.S. is grateful to Mitch Begelman and Mike Shull for discussions on various topics relevant to this work. Y.L. acknowledges support from the NSFC grants No. 12273031. The simulations were carried out at the National Supercomputer Center in Tianjin, using TianHe-1A. 
\end{acknowledgments}



\bibliography{ms}{} 
\bibliographystyle{aasjournal}




\label{lastpage}
\end{document}